\documentclass{PoS}

\title{Photon-jet correlations at RHIC and the LHC}

\ShortTitle{Photon-jet correlations at RHIC and the LHC}

\author{\speaker{Michael Klasen}
 \thanks{Work supported by the BMBF under contract 05H15PMCCA.}\\
 Institut f\"ur Theoretische Physik, Westf\"alische
 Wilhelms-Universit\"at M\"unster, Wilhelm-Klemm-Stra\ss{}e 9,
 D-48149 M\"unster, Germany\\
 E-mail: \email{michael.klasen@uni-muenster.de}}

\abstract{We present POWHEG predictions for photon-jet correlations at RHIC and the LHC.
 We show that the total transverse-momentum spectrum of photons and jets is modified not
 only by medium, but also by higher-order QCD effects, as is the distribution in their
 relative azimuthal angle. At the LHC, photon-jet measurements in the forward region
 allow to probe parton momentum fractions of lead ions down to 10$^{-4}$.}

\FullConference{EPS-HEP 2017, European Physical Society conference on High Energy Physics\\
		5-12 July 2017\\
		Venice, Italy}

\begin{document}

\section{Introduction}

Photon-jet correlations are important probes of the hot medium produced in heavy-ion
collisions \cite{Wang:1996yh}. In particular, the transverse momentum $p_T$ of the
photon, which does not
undergo strong interactions with the Quark-Gluon-Plasma (QGP), allows to calibrate the
$p_T$ of the jet at the time of its creation. A comparison of the observed jet-$p_T$
with the photon-$p_T$ then gives a quantitative measure of jet energy loss in the medium.
Similarly, the azimuthal angle difference $\Delta\phi$ between the jet and the photon
is modified by the QGP, whereas one expects both to be back-to-back {\em in vacuo}.
This simple picture is, however, based on perturbative QCD at leading order (LO) and,
as we will see in this contribution, must be modified in next-to-leading order (NLO)
and beyond.

Up to now, photon-jet correlations have been either analysed with LO Monte Carlo
generators like PYTHIA \cite{Skands:2014pea} or with NLO calculations like JETPHOX
\cite{Catani:2002ny}. While the former suffer from large theoretical scale
uncertainties, the latter have only up to three partons in the final state and thus
insufficient detail to describe all of the experimental observables. We have recently
combined NLO calculations with parton showers (PS) for photon production using the
POWHEG method \cite{Frixione:2007vw} and applied them to data from RHIC
\cite{Jezo:2016ypn} and the LHC \cite{Klasen:2017dsy}.

Typical results for photon-jet correlations at BNL's Relativistic Heavy-Ion Collider
RHIC are shown in Sec.\ \ref{sec:2}. In Sec.\ \ref{sec:3}, we demonstrate how
forward measurements of proton-lead collisions at CERN's Large Hadron Collider LHC
can give access to nuclear parton distribution functions (nPDFs) at very low values
of the parton momentum fraction $x$. We then summarise our results in Sec.\ \ref{sec:4}.

\section{Photon-jet correlations at RHIC}
\label{sec:2}

\begin{figure} 
\begin{center}
 \includegraphics[width=0.9\textwidth]{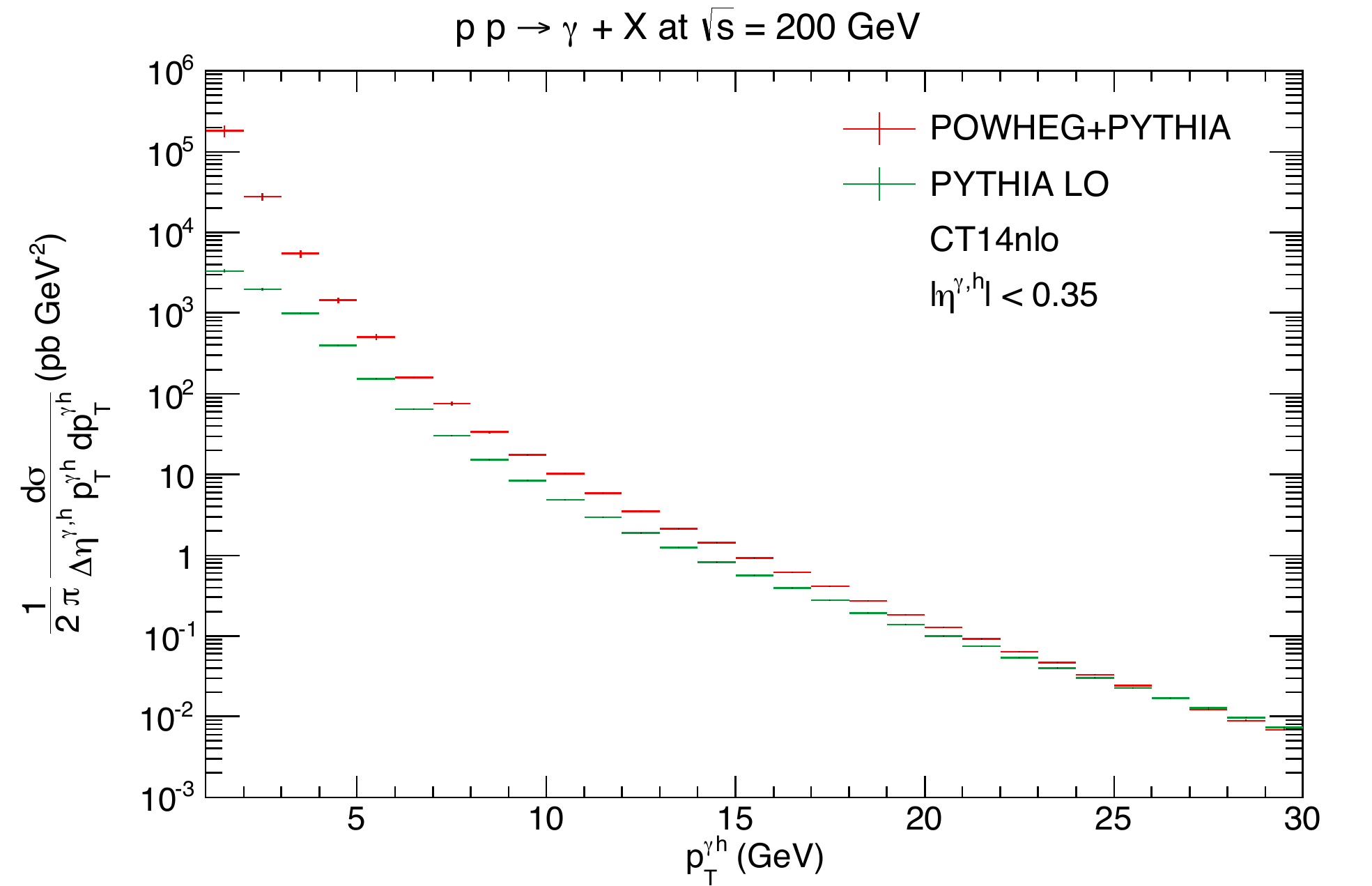}
 \caption{Transverse-momentum spectrum of the
 photon-hadron pair in pp collisions with $\sqrt{s}=200$ GeV at RHIC
 in LO+PS (green) and NLO+PS (red) \cite{Jezo:2016ypn}.}
 \label{fig:01}
\end{center}
\end{figure}

The PHENIX collaboration has measured photon-hadron jet correlations
in pp collisions at a centre-of-mass energy of $\sqrt{s}=200$ GeV
\cite{Adare:2009vd}. Photons and charged hadrons were detected at
central rapidity $|\eta^{\gamma,h}|<0.35$.
In Fig.\ \ref{fig:01} we show the distribution in the combined
photon-hadron tranverse momentum $p_T^{\gamma h}$. The non-perturbative
region is avoided by applying individual cuts on $p_T^{\gamma,h}>1$ GeV.
At sufficiently high net $p_T^{\gamma h}$, the NLO+PS corrections (red) to
the LO prediction (green) are small, since all relevant scales are large.
The interesting region used for jet calibration is the region of small
$p_T^{\gamma h}$. While the individual transverse momenta still allow
the application of perturbation theory, higher-order corrections become
large due to soft gluon emission and must be resummed to all orders.
This is done reliably either with analytic resummation methods or,
as is the case here, by combining the NLO calculation with a PS.

\begin{figure} 
\begin{center}
 \includegraphics[width=0.9\textwidth]{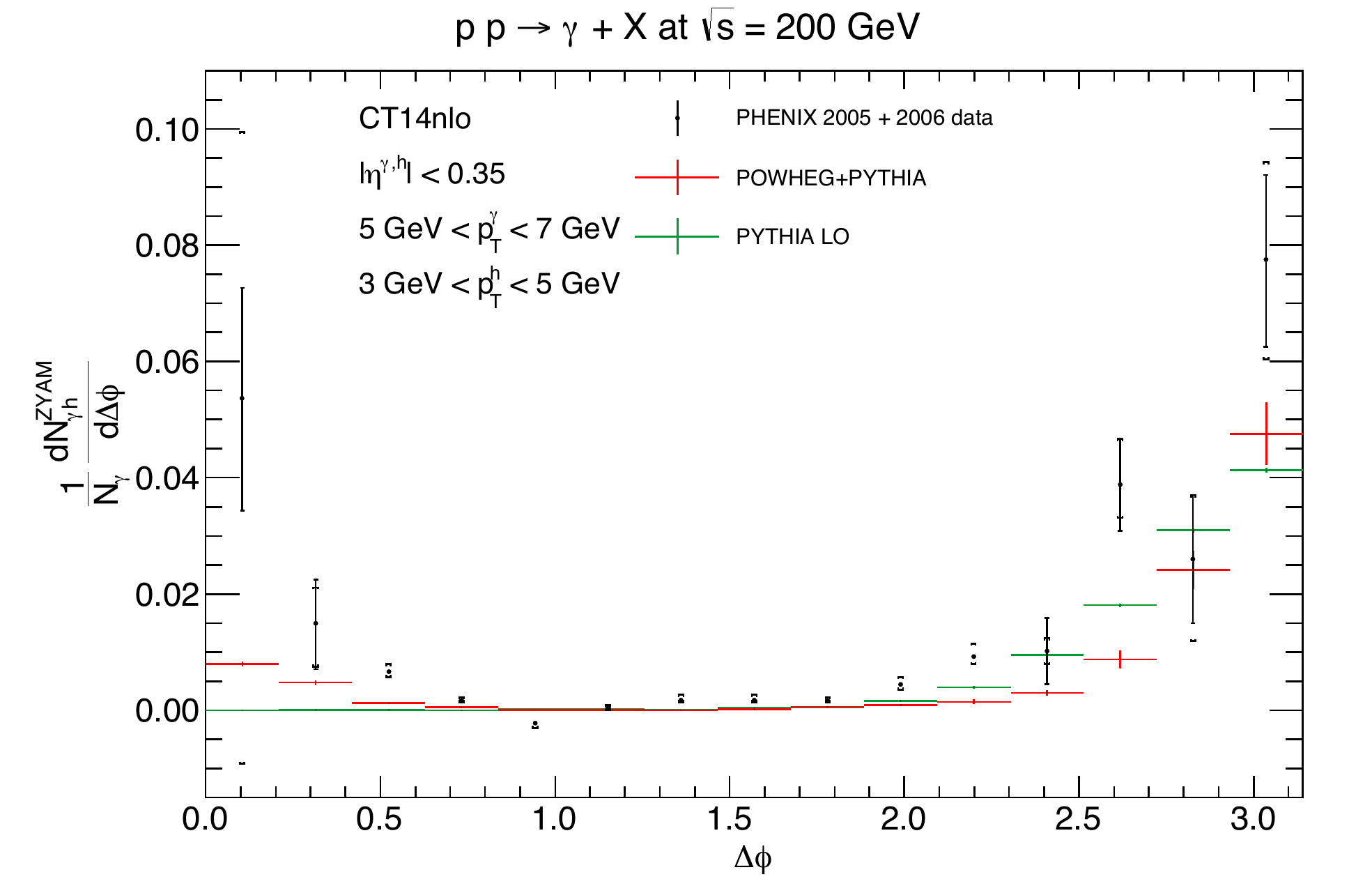}
 \caption{Azimuthal-angle correlation of the photon-hadron pair
 in pp collisions with $\sqrt{s}=200$ GeV at RHIC in LO+PS (green) and
 NLO+PS (red) and compared to PHENIX data (black) \cite{Adare:2009vd}.}
 \label{fig:02}
\end{center}
\end{figure}

To determine the azimuthal correlations, photons and jets were accepted
by PHENIX above transverse momenta $p_T^{\gamma,h}$ of 5 and 3 GeV, respectively.
The corresponding results are shown in Fig.\ \ref{fig:02}, subtracted
to exhibit Zero Yield at the Minimum (ZYAM). The away-side region with
photons and jets in opposite hemispheres ($\Delta\phi\sim\pi$) is reasonably
described already at LO (green). At NLO+PS (red), the rise in this region is
enhanced and closer to the data. More important is the near-side or fragmentation
region, where the LO predictions remain unrealistically flat, while those
at NLO+PS follow the rise of the data.

\section{Photon-jet correlations at the LHC}
\label{sec:3}

Photon-jet correlations can also help us to improve on our knowledge
of nuclear modifications of PDFs such as shadowing at low $x$.
The reason is that Deep-Inleastic Scattering (DIS) and Drell-Yan
(DY) processes constrain mostly quark and anti-quark nPDFs at intermediate
and large $x$, respectively \cite{Kovarik:2015cma}, while considerable
uncertainties remain in the nuclear gluon and sea quark PDFs at
small $x$. This could be changed by including in the global fits future
LHC pPb data from photon production in association with jets (or heavy quarks)
\cite{Stavreva:2010mw}.

\begin{figure}[ht]
\begin{center}
 \includegraphics[width=\textwidth]{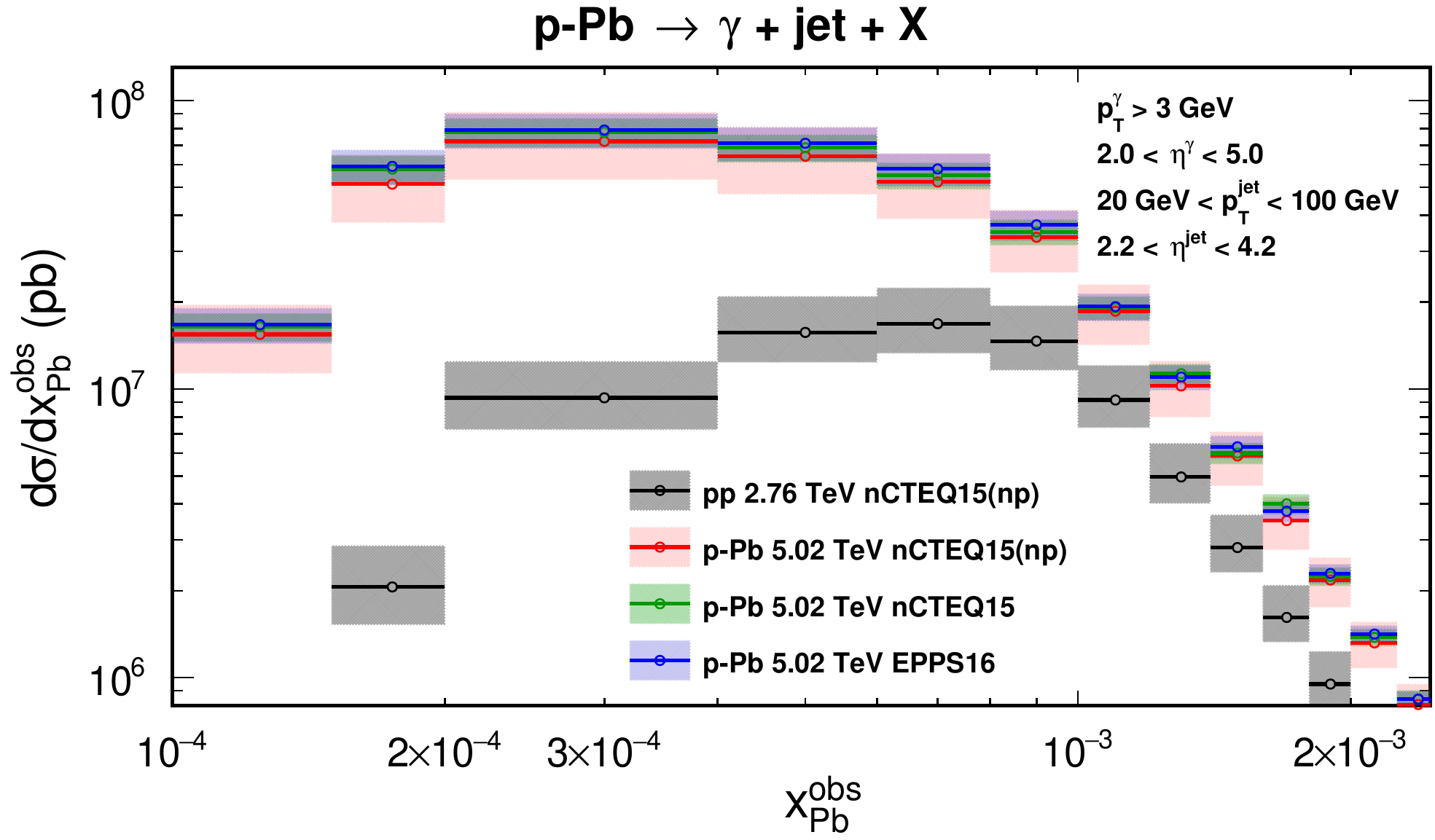}
 \caption{Differential distribution in the observed parton momentum fraction in
 lead ions in pPb collisions at the LHC \cite{Klasen:2017dsy}.}
 \label{fig:03}
\end{center}
\end{figure}

A useful variable to analyse photon-jet data directly in view of the nPDFs
is the observed parton momentum fraction
\begin{equation}
 x_{\rm Pb}^{\rm obs} := { p_T^\gamma e^{-\eta^\gamma} + p_T^{\rm jet} e^{-\eta^{\rm jet}}
 \over
 2 E_{\rm Pb}}
\end{equation}
in the lead ion. As one can see, small values of $x$ are best accessed
at small transverse momenta and/or forward rapidities.
Fig.\ \ref{fig:03} shows the differential cross section of photon-jet
production in pPb collisions at the LHC with a centre-of-mass energy
per nucleon of $\sqrt{s_{NN}}=5.02$ TeV. As expected, data from DIS and
DY alone (red) lead to large nPDF uncertainties at small $x$. The
addition of pion production data from RHIC (green) and weak-boson and jet
production data from LHC (blue) lead to similar improvements, which can also
be expected from photon-jet data. For comparison, predictions for pp
collisions at 2.76 TeV including theoretical scale uncertainties
are also shown (grey).

\section{Conclusions}
\label{sec:4}

In conclusion, we have computed photon-jet production at hadron colliders
at NLO and matched this calculation to PS using the POWHEG method. This
will allow in the future for predictions of photon and jet observables in
pp, pPb and PbPb collisions at RHIC or the LHC with reliable normalisation
and realistic detail of the final state.

Choosing three typical observables, we have demonstrated this assertion
for the transverse-momentum balance and azimuthal-angle correlation at RHIC
and for studies of nuclear PDF uncertainties at the LHC. The latter become
easily accessible if the data are directly analysed in terms of the observed
parton momentum fraction in the lead ion.

\section*{Acknowledgments}

We thank T.\ Jezo, C.\ Klein-B\"osing, F.\ K\"onig and H.\ Poppenborg for their collaboration
and M.\ Cacciari, D.\ d'Enterria and Y.J.\ Lee for useful discussions.

\end{document}